\documentclass[12pt]{iopart}

\usepackage{iopams}
\usepackage{graphicx}
\usepackage{color}

\begin{document}

\newcommand{\be}{\begin{equation}}
\newcommand{\ee}{\end{equation}}
\newcommand{\bea}{\begin{eqnarray}}
\newcommand{\eea}{\end{eqnarray}}
\newcommand{\nn}{\nonumber}
\newcommand{\de}{\partial}

% Greek alphabet
\def\a{\alpha}
\def\b{\beta}
\def\d{\delta}        \def\D{\Delta}
\def\e{\epsilon}
\def\eps{\varepsilon}
\def\f{\phi}          \def\F{\Phi}
\def\vf{\varphi}
\def\g{\gamma}        \def\G{\Gamma}
\def\h{\eta}
\def\i{\iota}
\def\j{\psi}          \def\J{\Psi}
\def\k{\kappa}
\def\lam{\lambda}   \def\L{\Lambda}
\def\m{\mu}
\def\n{\nu}
\def\o{\omega}   \def\O{\Omega}
\def\p{\pi}      \def\P{\Pi}
\def\q{\theta}   \def\Q{\Theta}
\def\r{\rho}
\def\s{\sigma}   \def\S{\Sigma}
\def\t{\tau}
\def\u{\upsilon}  \def\U{\Upsilon}
\def\x{\xi}      \def\X{\Xi}
\def\z{\zeta}

\title[Electrons--O$_2$ RVE cross sections]{Resonant vibrational-excitation cross sections and rate constants for low-energy electron scattering by molecular oxygen}

\author{
V. Laporta$^{1,3,*}$\footnote[0]{$^*$ v.laporta@ucl.ac.uk},
R. Celiberto$^{2,3}$
and J. Tennyson$^{1}$
}

\address{$^1$ Department of Physics and Astronomy, University College London, London WC1E 6BT, UK}
\address{$^2$ Dipartimento di Ingegneria Civile, Ambientale, del Territorio, Edile e di Chimica, Politecnico di Bari, 70125 Bari, Italy}
\address{$^3$ Istituto di Metodologie Inorganiche e dei Plasmi - sez. di Bari, CNR, 70125 Bari, Italy}

\begin{abstract}
Resonant vibrational-excitation cross sections and rate constants for electron scattering by molecular oxygen are presented. Transitions between all 42 vibrational levels of O$_2(\textrm{X}\ ^3\Sigma_g^- $) are considered. Molecular rotations are parameterized by the rotational quantum number $J$ which is considered in the range 1 to 151. The lowest four resonant states of O$_2^-$, $^2\Pi_g$, $^2\Pi_u$, $^4\Sigma_u^-$ and $^2\Sigma_u^-$, are taken into account. The calculations are performed using the fixed-nuclei R-matrix approach to determine the resonance positions and widths, and the boomerang model to characterize the nuclei motion. Two energy regions below and above 4~eV are investigated: the first one is characterized by sharp structures in the cross section, and the second by a broad resonance peaked at 10~eV. The computed cross sections are compared with theoretical and experimental results available in literature for both the energy regions, and are made available for use by modelers. The effect of including rotational motion is found to be non-negligible.
\end{abstract}

\maketitle

\section{Introduction \label{sec:intro}}
Low-energy electron scattering by molecular oxygen is an important process for a wide class of natural and technological systems. Gaseous discharges, laboratory and astrophysical plasmas, combustion chemistry and biological science provide some examples where electron--O$_2$ scattering data are required~\cite{PhysRevLett.76.3534, PhysRevLett.31.969, allan_0953-4075-28-23-021, 0953-4075-27-14-048, PhysRevA.48.1214, 1367-2630-5-1-114, CTPP:CTPP200710073}. In particular these collisions play an important role in re-entry physics, and in the chemistry of the Earth's atmosphere where oxygen is the second most abundant species and where vibrationally excited molecules gives a fundamental contribution in redistributing the energy released into the atmospheric plasmas~\cite{bultel:043502}.

Resonant vibrational-excitation (RVE) is among the most efficient ways to populate the high vibrational levels of molecules in those plasmas where non-equilibrium conditions are present. In fact, direct vibrational-excitation of homonuclear diatomic molecules by electron impact is, in general, an inefficient process because of both the small electron-to-molecule mass ratio and the lack of an electric dipole moment. On the other hand, at energies where the incident electron can attach to a molecule and form a temporarily resonant anionic state, vibrational transition probabilities can be enhanced by orders of magnitude.

Complete sets of cross sections and rate coefficients for RVE have been obtained previously for H$_2$, CO, N$_2$ and NO~\cite{PhysRevA.77.012714, 0963-0252-21-4-045005, 0963-0252-21-5-055018} which are major components of the terrestrial or planetary atmospheres. In the present paper we extend the calculations to the resonant vibrationally-resolved cross sections and rate coefficients for electron-impact excitation of the O$_2$ molecule in its electronic ground state. These data represent the basic information required to develop molecular collisional-radiative models based on a state-to-state approach~\cite{bultel:043502}.

Low-energy electron--O$_2$ resonant scattering involves four resonant electronic states of the O$_2^-$ ion, according to the following process:
\be e + \textrm{O}_2(\textrm{X}\ ^3\Sigma^-_g; v) \longrightarrow \textrm{O}^-_2(^2\Pi_g, ^2\Pi_u, ^4\Sigma^-_u, ^2\Sigma^-_u) \longrightarrow e + \textrm{O}_2(\textrm{X}\ ^3\Sigma^-_g; v')\,, \label{eq:O2reaction}\ee
where $v$ and $v'$ represent, respectively, the initial and final vibrational levels. Previous theoretical and experimental studies of RVE of O$_2$ show, as discussed below, that below about 4~eV the dynamics are dominated by the $^2\Pi_g$ resonant state of O$_2^-$ and the cross sections consist of a set of very sharp peaks; above 4~eV, in the energy region around 10~eV, a broad peak is observed which is mainly determined by the $^4\Sigma_u^-$ resonant state with a non-negligible contribution coming from the $^2\Sigma_u^-$ state. In this paper both energy regimes are investigated.

The plan of the paper is as follows: Section~\ref{sec:THmodel} summarizes the basic equations of the theoretical model used to treat the nuclear dynamics; Section~\ref{sec:PEC} describes the fixed-nuclei O$_2$ and O$_2^-$ potentials energy curves; while Section~\ref{sec:results} presents and discusses our results which are also compared  with those from previous theoretical studies and experimental measurements.

\section{Theoretical model \label{sec:THmodel}}
In general a resonant state is a quasi-bound state that decays with a finite lifetime. This is the case of resonant electron--molecule scattering, where the incoming electron, temporarily trapped by the target molecule, is then re-emitted. This process may occur with the concomitant vibrational excitation of the neutral molecule. Since the typical vibrational time of the target nuclei is comparable with resonant lifetimes~\cite{PhysRev.125.229}, the response of the nuclei needs to be taken into account during an electronic collision. Our theoretical approach to treat the nuclear motion, briefly described below, is characterized by the well-known `boomerang model'~\cite{PhysRevA.20.194}, which has been recently successfully applied to a number of resonant collisions problems~\cite{PhysRevA.77.012714, 0963-0252-21-4-045005, 0963-0252-21-5-055018}. This model, which is  derived from the more general nonlocal-complex-potential theory, describes the nuclear dynamics within the Born-Oppenheimer approximation, once the complex electronic potential energy is provided. A complete description of this model and its limits can be found in the papers~\cite{PhysRevA.77.012714, 0963-0252-21-4-045005, 0963-0252-21-5-055018} and references therein. The boomerang model approximation is based on two requirements~\cite{PhysRevA.23.1089, PhysRevA.77.012710}: (i) that the electron energy $\e$ be greater than the vibrational levels spacing of the neutral target,
\be |\e_v-\e_{v'}|\ll\e\,,\hspace{3cm}\forall\, v,v'\,, \label{eq:spacing_approx}\ee and (ii) that the electron energy is around the resonance position $\e_d$, \be \e\approx\e_d(R)\,. \label{eq:respos_approx}\ee
By means of the \emph{ansatz} in Eqs.~(\ref{eq:spacing_approx}) and (\ref{eq:respos_approx}) the equations in the nonlocal theory become local and energy-independent. Since electron--O$_2$ scattering considered here includes many resonant states, a multi-resonance extension of the boomerang model is required.

Referring to the process in (\ref{eq:O2reaction}) the total conserved energy of the system is $E = \epsilon + \epsilon_v$, where $\epsilon$ is the incoming electron energy and $\epsilon_v$ is the energy of initial O$_2$ vibrational level. Labeling the four O$_2^-$ symmetries as $i=1\ldots4=\{$$^2\Pi_g$, $^2\Pi_u$, $^4\S^-_u$, $^2\S^-_u\}$, the equation that governs the nuclear dynamics of the resonant states can be cast in the following vectorial form: \be \left( E-\hat H \right)\vec \xi(R) = \vec{\mathcal V}^{\textrm{in}}\,, \label{eq:nucl_dyn} \ee for each of the corresponding four resonant nuclear wave functions $\xi_i(R)$. The diagonal elements of the effective Hamiltonian $\hat H$ are written in terms of the nuclear kinetic operator $T_N$, which in turn depends on the rotational quantum number $J$, and the complex resonant potentials,
\be
H_{ii} = T_N + V^-_i - \frac i2 \Gamma_i\,,\hspace{1cm}i=1\ldots4\,. \label{eq:Hii}
\ee
In the boomerang model the width functions $\Gamma_i$ in Eq.~(\ref{eq:Hii}) are considered energy-independent quantities. Conversely the off-diagonal elements of the Hamiltonian $\hat H$ in principle contain direct and indirect couplings between the resonant states~\cite{estrada:152}. In the case at hand, due to the different symmetries of the states, these two couplings are suppressed and the four resonances can be treated as non-interacting and therefore independent of each other. In Eq.~(\ref{eq:nucl_dyn}), $\vec{\mathcal V}^{\textrm{in}}$ represents the entry-amplitude defined by: \be {\mathcal V}^{\textrm{in}}_i = \sqrt{\frac{1}{2\pi}\frac{\Gamma_i}{k}}\,\chi_v\,, \label{eq:entry_amplitude} \ee where $k$ is the  momentum of the incoming electron and $\chi_v(R)$ is the wave function corresponding to the initial vibrational level $v$ of O$_2$ potential, solution of the wave equation: \be (T_N+V_0)\chi_v=\e_v\chi_v\,. \label{eq:neutral_weq} \ee The neutral molecule potential $V_0$ in Eq.~(\ref{eq:neutral_weq}) and the complex potentials in Eq.~(\ref{eq:Hii}) are discussed in the next Section.

One can derive the $T$-matrix elements from Eq.~(\ref{eq:nucl_dyn}) as:
\bea
T_{ij} &=& \langle \mathcal{V}_i^{\textrm{out}}|\xi_j \rangle \nn
\\
&=& \sum_{n=1}^4\langle \mathcal V_i^{\textrm{out}}(\hat H-E)_{jn}^{-1}\mathcal V_n^{\textrm{in}} \rangle\,,\label{eq:Tmatrixel}
\eea
where $\langle\cdots\rangle$ means integration over internuclear distance $R$ and $\mathcal V_i^{\textrm{out}}$ is the exit-amplitude written, analogously to the Eq.~(\ref{eq:entry_amplitude}), in terms of the wave function $\chi_{v'}(R)$ of the $v'$-th final O$_2$ vibrational level and the outgoing electron momentum $k'$. It can be demonstrated that the diagonal matrix elements of the $T$-matrix couple totally symmetric modes and the off-diagonal elements couple non-totally symmetric modes. Only totally symmetric modes are allowed for the excitation from electronic ground state of O$_2$~\cite{estrada:152}. It is well-known that for very low-energies the local model can fails in describing accurately the scattering process. In order to take into account the threshold effects, in particular when the conditions in Eq.~(\ref{eq:spacing_approx}) and (\ref{eq:respos_approx}) are not fully respected, a barrier penetration factor, represented by an \emph{ad-hoc} energy-dependent function can be added in the definition of the entry- and exit-amplitude~\cite{PhysRevA.71.052714}. To test the effect of the penetration factor on the cross sections, calculations were performed assuming entry-amplitude has the form adopted by Trevisan \emph{et al.}~\cite{PhysRevA.71.052714}. This only changed the results by about 10 \%\ compared to the cross sections calculated with the local model;  in some case, the comparison with the experiment was  worse. These results implies that the penetration factor, in the analytical form reported in Ref.~\cite{PhysRevA.71.052714}, is not crucial in modulating the cross sections for O$_2$ molecule.

Neglecting  the interference terms between resonances in Eq.~(\ref{eq:Tmatrixel}), which is a consequence of our assumption that the off-diagonal elements of $\hat H$ are vanishing small, the final RVE cross section can be written as a superposition of four independent contributions:
\be
\sigma_{vv'}(\epsilon) = \frac{16 \pi^4 m_e}{\hslash^2} \frac{k}{k'} \sum_{i=1}^4 g_i\left| T_{ii} \right|^2\,, \label{eq:cs}
\ee
where $g_i$ is the spin-statistical factor for the $i^{th}$ resonant state. Once the cross sections are provided, the rate constants for process (\ref{eq:O2reaction}) can be obtained. Assuming a Maxwellian electron energy distribution, at temperature $T_e$(eV), the corresponding rate constant, $K_{vv'}(T_e)$, reads as: \be K_{vv'}(T_e) = \sqrt\frac8{m_e\pi} \left( \frac1{T_e} \right)^{3/2} \int_{\epsilon_{th}}^\infty d\e\,\e\,\s_{vv'}(\e)\,e^{-\frac\e{T_e}}\,, \label{eq:ratec} \ee where $\e_{th}=\e_{v'}-\e_v$ is the threshold energy of the process.

In the calculations below we only consider the $^{16}\textrm{O}$ oxygen isotope. Since $^{16}$O has zero total nuclear spin, $^{16}$O must have a symmetric total wave function. From Raman spectroscopy experiments~\cite{PhysRev.97.937}, it can be deduced that electronic ground state of O$_2$ is $^3\Sigma_{g}^-$, which is antisymmetric. Thus only odd values for rotational quantum number $J$ are allowed. Therefore the lowest rotational level in the ground state of O$_2$ molecule is $J=1$ and $\Delta J$ must be even.

\section{O$_2$ and O$_2^-$ potential energy curves \label{sec:PEC}}
The potential energy curve for the ground state $\textrm{X}\ ^3\Sigma^-_g$ of O$_2$ was calculated using the \emph{ab initio} quantum chemical code MOLPRO~\cite{MOLPRO_brief} within the Multi-Reference Configuration-Interaction (MRCI) model and using an aug-cc-pVQZ basis. The resulting potential curve supports 42 vibrational levels. Some spectroscopic parameters are provided in Table~\ref{tab:PESparam} and compared with available data values. The numerically calculated vibrational level energies are shown in Table~\ref{tab:O2viblev}.
\begin{table}[t]
\caption{Reduced mass ($\mu$), dissociation energy ($D_e$) and equilibrium distance ($R_e$) for O$_2$ and O$_2^-$ potentials. Electron affinity (eA) of O$_2$ and the crossing point ($R_c$) between  the O$_2$ and O$_2^-$ potential energy curves are also given. Literature values, where available, are given in parenthesis. \label{tab:PESparam}}
\begin{indented}
\item[]
\begin{tabular}{c@{\,}lccccc}
  \hline\hline
  &&  O$_2$& \multicolumn{4}{c}{
  \begin{tabular}{c}
  ~~~~~~~~~~~~~~~O$_2^-$~~~~~~~~~~~~~~~\\\hline
  \end{tabular}
  }
  \\
  && X $^3\Sigma^-_g$ & $^2\Pi_g$ & $^2\Pi_u$ & $^4\S^-_u$ & $^2\S^-_u$ \\
  \hline
  $\mu$ & (a.u.) & 14582.6 \\
  $D_e$ & (eV)   & 5.10 (5.12~\cite{bytautas:074109}) & 4.02 & 0.83 & 1.54 & 0.73 \\
  $R_e$ & (a.u.) & 2.29 (2.28~\cite{bytautas:074109}) & 2.55 & 3.38 & 3.47 & 3.73 \\
  $R_c$ & (a.u.) & --    & 2.34 & 3.20 & 3.03 & 3.25 \\
%  eA    & (eV)   &  --   & \multicolumn{4}{c}{1.45 (1.46~\cite{PhysRevA.81.022503})} \\
  eA    & (eV)   &  1.45 (1.46~\cite{PhysRevA.81.022503}) \\
  \hline\hline
\end{tabular}
\end{indented}
\end{table}

\begin{table}[t]
\caption{Calculated vibrational levels of O$_2(\textrm{X}\ ^3\Sigma^-_g)$. Energies are given in eV. \label{tab:O2viblev}}
\begin{indented}
\item[]
\begin{tabular}{cc@{\hspace{25pt}}cc@{\hspace{25pt}}cc}
 \hline \hline
 $v$ & $\e_v$ & $v$ & $\e_v$ & $v$ & $\e_v$  \\
 \hline
 0 &   0.000  &    14 &   2.435  &    28 &   4.280\\
 1 &   0.196  &    15 &   2.587  &    29 &   4.382\\
 2 &   0.388  &    16 &   2.735  &    30 &   4.476\\
 3 &   0.573  &    17 &   2.881  &    31 &   4.565\\
 4 &   0.756  &    18 &   3.024  &    32 &   4.651\\
 5 &   0.937  &    19 &   3.164  &    33 &   4.730\\
 6 &   1.117  &    20 &   3.301  &    34 &   4.794\\
 7 &   1.291  &    21 &   3.436  &    35 &   4.847\\
 8 &   1.461  &    22 &   3.568  &    36 &   4.898\\
 9 &   1.629  &    23 &   3.696  &    37 &   4.938\\
10 &   1.796  &    24 &   3.821  &    38 &   4.960\\
11 &   1.960  &    25 &   3.942  &    39 &   4.976\\
12 &   2.122  &    26 &   4.059  &    40 &   4.987\\
13 &   2.281  &    27 &   4.172  &    41 &   4.994\\
 \hline \hline
\end{tabular}
\end{indented}
\end{table}

\begin{figure}
\begin{indented}
\item[]
\begin{tabular}{cc}
\includegraphics[scale=.7,angle=0]{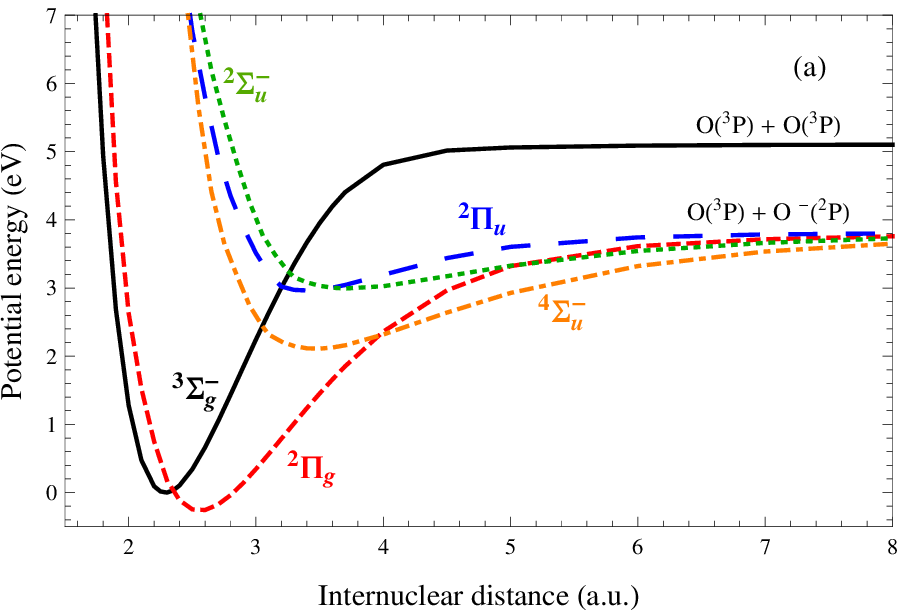}&
\includegraphics[scale=.7,angle=0]{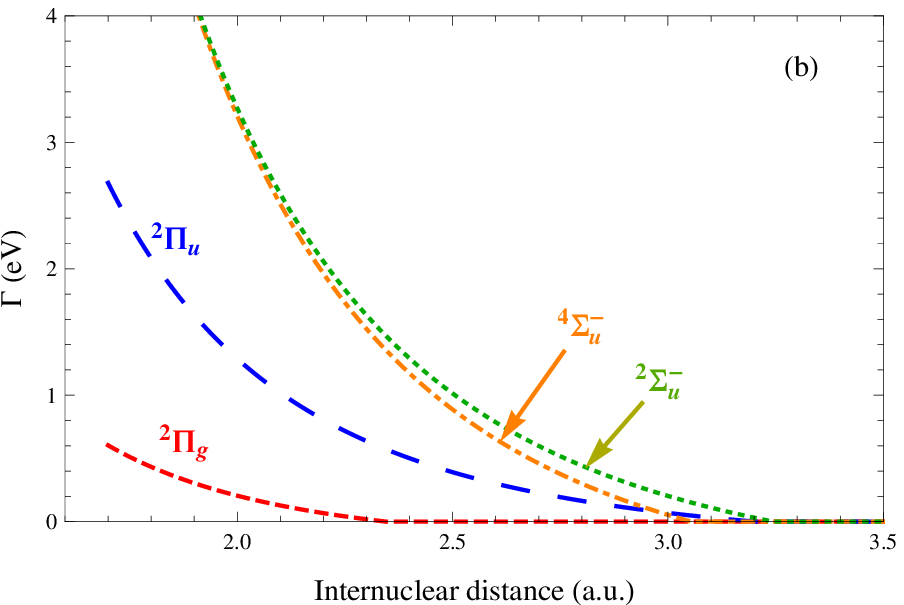}
\end{tabular}
\end{indented}
\caption{Potential energy curves (a) and resonance widths (b) as a function of internuclear distance for the lowest four states of O$_2^-$: $^2\Pi_g$ (short-dashed line), $^2\Pi_u$ (long-dashed line), $^4\S^-_u$ (dot-dashed line) and $^2\S^-_u$ (dotted line). The solid curve gives the
corresponding potential energy curve for the X$\ ^3\S^-_g$ of O$_2$ ground state. \label{fig:O2pes}}
\end{figure}

The potential energy curves  for the lowest four states of O$_2^-$ were calculated at internuclear distances greater than the crossing point, $R_c$, using MOLPRO and the same model and basis used for O$_2$. At these internuclear separations the anion potentials are real and the O$_2^-$ becomes stable~\cite{Stibbe1999532}. At shorter internuclear separations, where the O$_2^-$ behaves as a resonance and the potential energy curves are complex, resonance positions and widths were taken from the theoretical results of Noble \emph{et al.}~\cite{PhysRevLett.76.3534} who used a fixed-nuclei R-Matrix approach~\cite{Tennyson201029}. Since Noble \emph{et al.} were interested only in the first vibrational levels of O$_2$, the potential energy curve for neutral oxygen used in their calculations is not very accurate and it is significantly different from that used in the present paper. In practice we find that the final cross sections are very sensitive to the resonant potentials, in particular for high vibrational levels. For these reasons the resonance positions for the four symmetries were taken from Noble \emph{et al.} and added to our calculated O$_2$ energies, generating the corresponding resonance curves. These curves were then joined to our O$_2^-$ MOLPRO results. The final potential energies, used in our cross section calculations, are shown in Fig.~\ref{fig:O2pes} and some characteristic parameters are collected in Table~\ref{tab:PESparam}. These curves give an electron affinity for O$_2$ of 1.45~eV, which is compatible with the high accuracy first-principles theoretical calculations~\cite{PhysRevA.81.022503} and the experimental value of 1.46~eV~\cite{springerlink:10.1140/epjd/e2005-00069-9}.

\section{Results and discussion \label{sec:results}}

To validate the calculated cross sections using the model presented in the previous sections comparisons are made with previous studies. In particular the theoretical and experimental data of Noble \emph{et al.}~\cite{PhysRevLett.76.3534}, the measurements of Wong \emph{et al.}~\cite{PhysRevLett.31.969} and those of Allan~\cite{allan_0953-4075-28-23-021} were used. Where not explicitly specified, the RVE cross section is understood to be the sum over the four resonant states of O$_2^-$, as given in Eq.~(\ref{eq:cs}), and calculated for $J=1$.  As discussed above for electron--O$_2$ resonant scattering, the RVE cross section can be divided into two distinct energy regions: below 4~eV, where the cross section exhibits sharp structures, and above this energy, where a broad maximum, peaked around 10~eV, is observed.

\begin{figure}[t]
\begin{indented}
\item[]
\begin{tabular}{cc}
\includegraphics[scale=.7,angle=0]{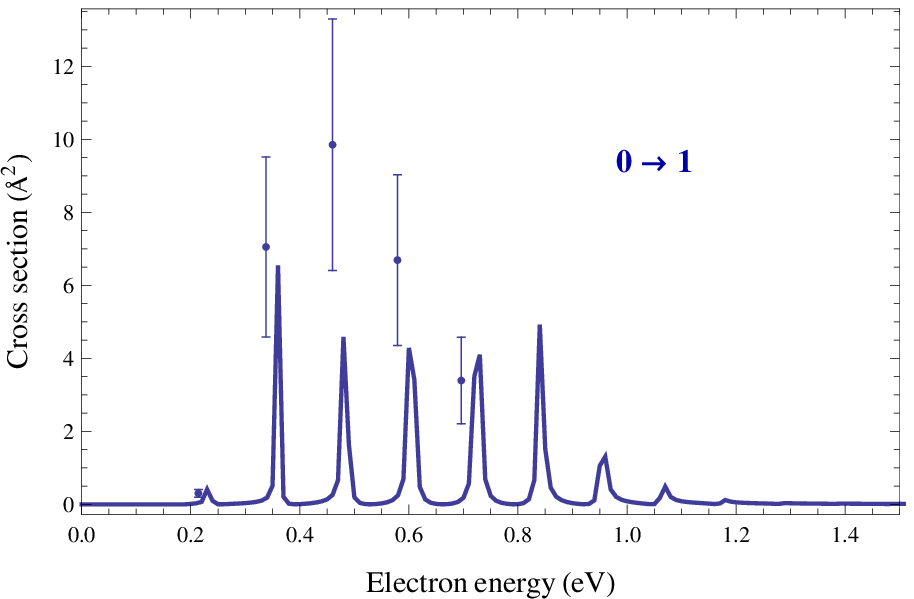}&
\includegraphics[scale=.7,angle=0]{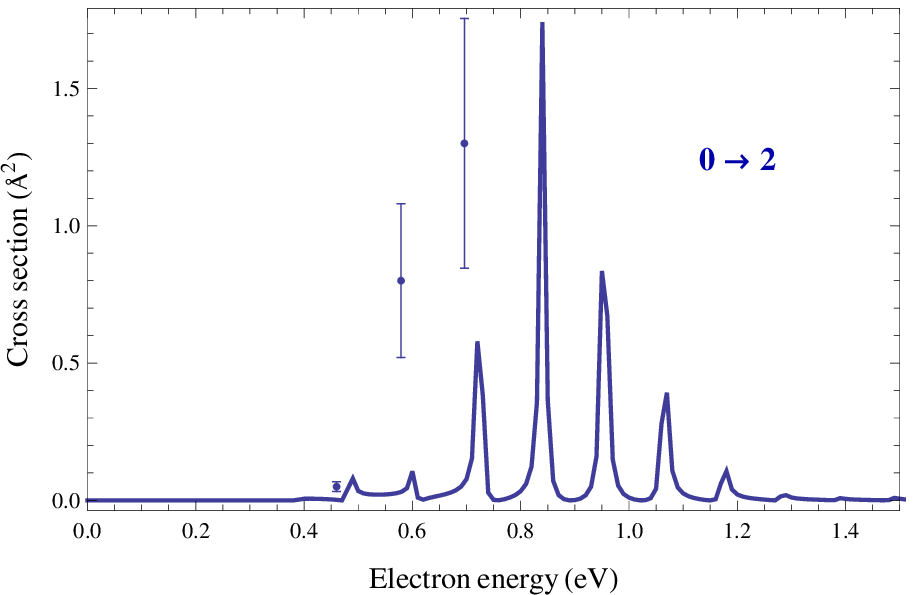}
\end{tabular}
\end{indented}
\caption{Electron--O$_2$ resonant vibrational excitation cross sections, at low energy where $^2\Pi_g$ state dominates, for $0\to1$ and $0\to2$ transitions compared with experimental results~\cite{allan_0953-4075-28-23-021, itikawa:1}. \label{fig:cfr_allan}}
\end{figure}

Figure~\ref{fig:cfr_allan} shows the calculated RVE cross sections compared with Allan's results~\cite{allan_0953-4075-28-23-021}, for the first vibrational levels, as given by Itikawa~\cite{itikawa:1}. In his paper Allan reports energy-integrated cross sections that, in order to extract the absolute values, Itikawa divides by the resonance width. The figure shows Itikawa's results. At energy below 2~eV the $^2\Pi_g$ state of the O$_2^-$ dominates and since this is the most stable anion state (see the lowest curve in Fig.~\ref{fig:O2pes}(a)) and the longest lived (see the narrow resonance width in Fig.~\ref{fig:O2pes}(b)), the cross section consists of a series of narrow spikes. As noted by Allan, the narrowness of the peaks makes experimental measurements difficult at low energy near the threshold, where the resonance width is comparable to the instrumental resolution. As a consequence the experimental error is large, about $35\%$. Within this experimental uncertainty, our absolute cross sections are generally lower than the experimental ones, showing a discrepancy for the highest peaks no larger than about a factor of two. The peak positions of the present results, reflecting the energy of the resonant vibrational levels, are slightly shifted with respect the experimental points.

\begin{figure}[t]
\begin{indented}
\item[]
\begin{tabular}{cc}
\includegraphics[scale=.7,angle=0]{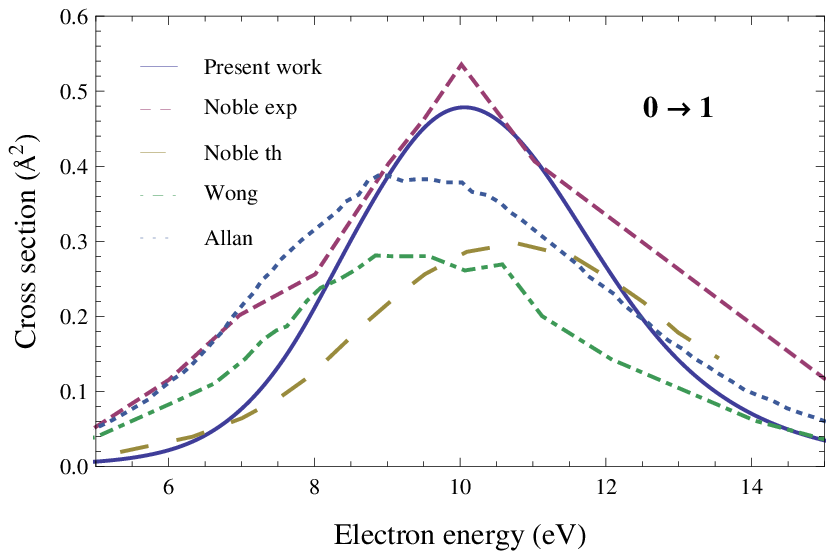} &
\includegraphics[scale=.7,angle=0]{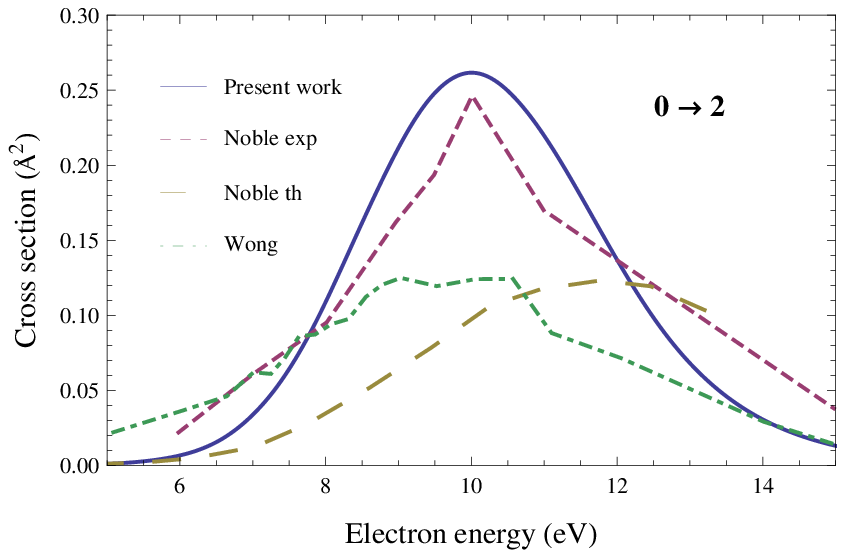}
\\
\includegraphics[scale=.7,angle=0]{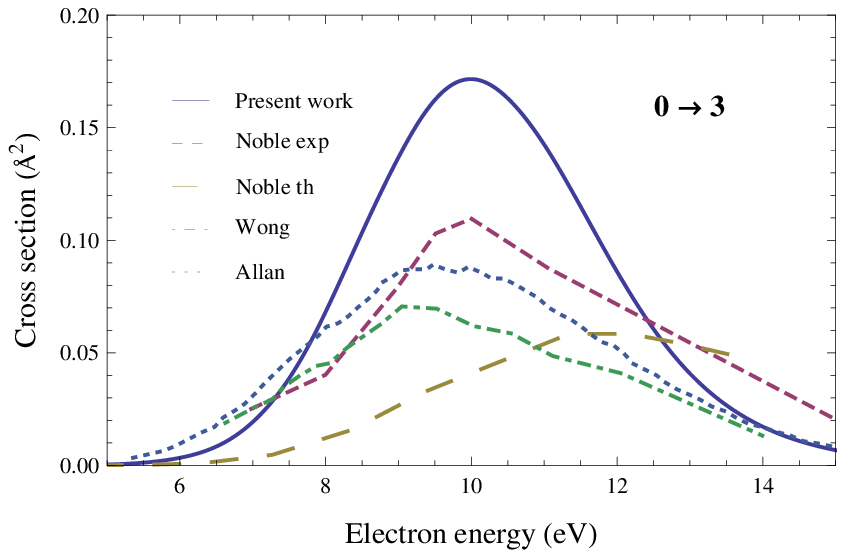} &
\includegraphics[scale=.7,angle=0]{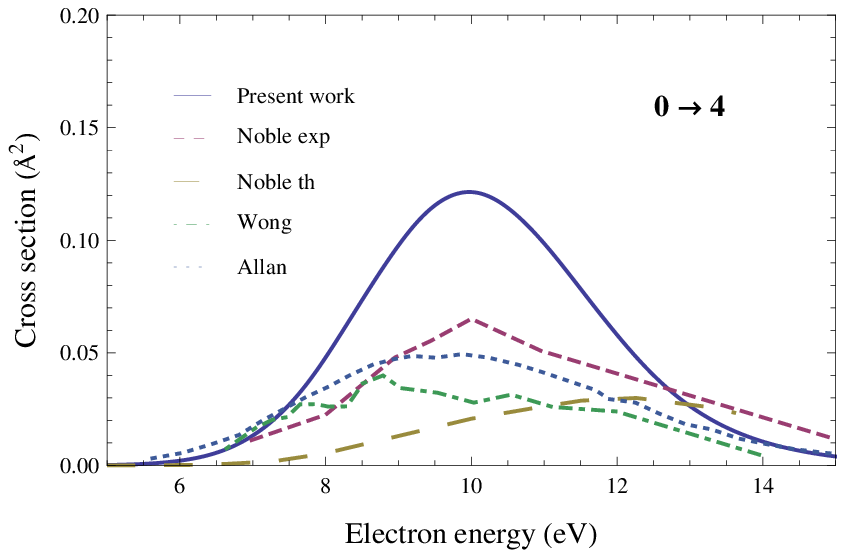}
\end{tabular}
\end{indented}
\caption{Angular integrated electron--O$_2$ vibrational excitation cross section in the 10~eV region. Our calculated cross sections (solid curve) are compared with experimental (dashed line, error of 20$\%$--26$\%$) and theoretical (long-dashed line) results of Noble \emph{et al.}~\cite{PhysRevLett.76.3534}, Wong \emph{et al.}'s measurements (dot-dashed line, error about 30$\%$)~\cite{PhysRevLett.31.969} and Allan's experiment (dotted line, error 35$\%$)~\cite{allan_0953-4075-28-23-021}.\label{fig:cfr_all}}
\end{figure}

\begin{figure}[t]
\begin{indented}
\item[]
\begin{tabular}{c}
\includegraphics[scale=.7,angle=0]{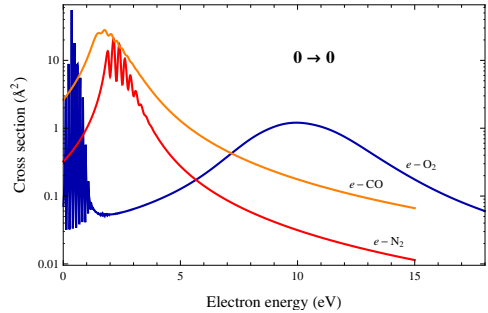}
\end{tabular}
\end{indented}
\caption{Comparison of electron--O$_2$, --CO~\cite{0963-0252-21-4-045005} and --N$_2$~\cite{0963-0252-21-5-055018} cross sections at 10~eV. \label{fig:cfr_O2-N2-CO}}
\end{figure}

Figure~\ref{fig:cfr_all} shows the behavior of the cross sections in the second region where the $^4\Sigma^-_u$ resonance is supposed to be dominant. The principal feature is a  broad peak in the cross section, located around 10~eV, which contrasts with the sharp shape of the peaks present at lower energies. Since a 10~eV electron has sufficient energy to dissociate the O$_2$ molecule, the interference between the neutral vibrational wave function in the entry-amplitude of Eq.~(\ref{eq:nucl_dyn}) with the continuum part of the anionic spectrum, gives rise to a smooth shape for the cross section. This behavior is not present in other similar systems such as electron--CO~\cite{0963-0252-21-4-045005} or electron--N$_2$~\cite{0963-0252-21-5-055018}, where, above the dissociation threshold, the cross section shows a continuum unstructured shape which decreases monotonically with the energy. A quantitative comparison is given in Fig.~\ref{fig:cfr_O2-N2-CO} for the transition $0\to0$. This behavior arises from the different relative positions of the O$_2(\textrm{X}\ ^3\Sigma^-_g)$ ground state  and the $^4\Sigma^-_u$ resonant state potential energy curves compared to the relative positions of the curves for N$_2$ and N$_2^-$ or for CO and CO$^-$~\cite{O2continuum}.

\begin{figure}[t]
\begin{indented}
\item[]
\begin{tabular}{cc}
\includegraphics[scale=.7,angle=0]{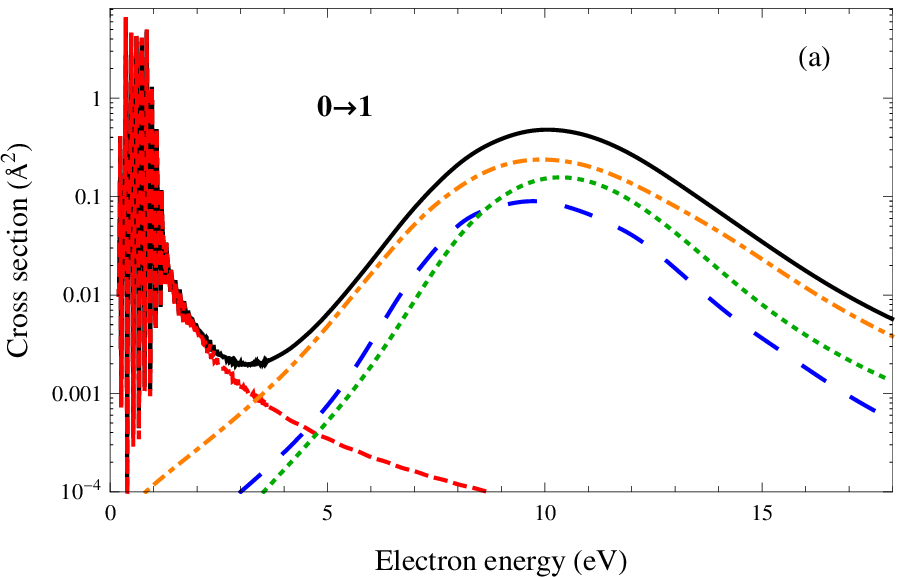}&
\includegraphics[scale=.7,angle=0]{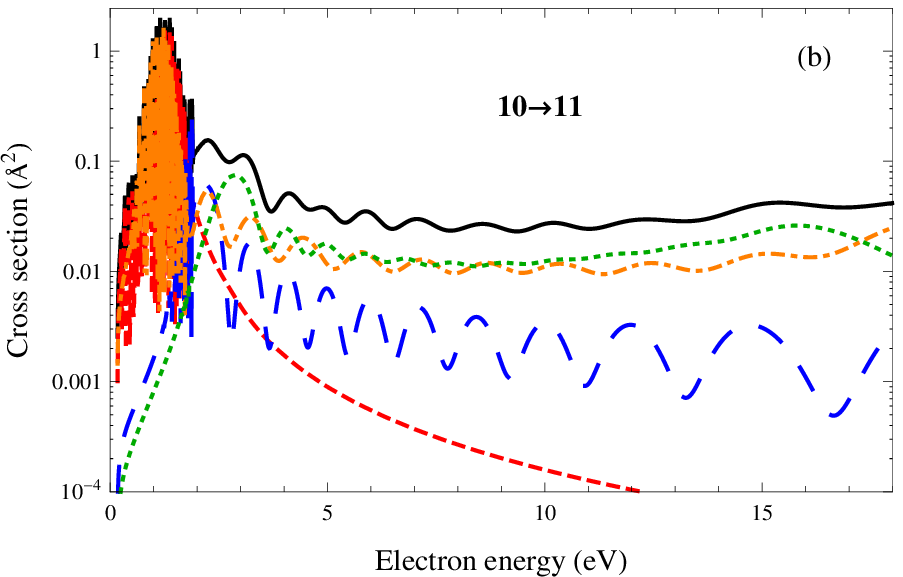}
\end{tabular}
\end{indented}
\caption{Electron--O$_2$ total cross sections (solid curve) separated into the four contributions coming from O$_2^-$ states: $^2\Pi_g$ (short-dashed line), $^2\Pi_u$ (long-dashed line), $^4\S^-_u$ (dot-dashed line) and $^2\S^-_u$ (dotted line) for two examples of a $\Delta v =1$ vibrational excitation. \label{fig:allsymmxsec}}
\end{figure}

Figure~\ref{fig:cfr_all} compares our results with previous studies~\cite{PhysRevLett.76.3534, PhysRevLett.31.969, allan_0953-4075-28-23-021}. The agreement is in very satisfactory agreement, in particular for the $0\to 1$ and $0\to 2$ transitions, with Noble \emph{et al.}'s experiments~\cite{PhysRevLett.76.3534} where both the maximum of the feature and its width coincide. Some disagreement is observed with the theoretical results of Noble \emph{et al.} (long-dashed curves) which however included, in their calculations, only the $^4\Sigma^-_u$ symmetry: this explains the discrepancies with the present results which consider all the four symmetries of O$_2^-$. In fact, as can be seen from Fig.~\ref{fig:allsymmxsec}(a), where the individual contributions to the cross section are shown, the main contribution at 10~eV comes from the $^4\Sigma_u^-$ symmetry but the contributions from the $^2\Sigma_u^-$ and $^2\Pi_u$ are not completely negligible.

Figure~\ref{fig:allsymmxsec}(a) also shows the dominance, in the total cross section, of the $^2\Pi_g$  symmetry at low energies. This dominance is true for low vibrational levels, such as the transition $0\to1$. For transitions involving higher vibrational levels the behavior is quite different. In fact, as the starting vibrational level approaches the bottom of a higher anionic potential curves, the discrete structure of the corresponding resonant vibrational levels become evident and new peaks appear in the cross section. This is the case, for example, for the $10\to11$ transition shown in Fig.~\ref{fig:allsymmxsec}(b), where the cross section near  threshold is dominated by the peaks due to the $^2\Pi_g$ and $^4\Sigma^-_u$ symmetries. Fig.~\ref{fig:allsymmxsec}(b) also shows oscillations in the cross section for energies above 2~eV that substitute the single broad peak present for the transitions $0\to v'$. This behavior is probably due to the interaction between the  neutral wave function of the vibrational level 10 and the continuum spectrum of the O$_2^-$ ion~\cite{Celiberto2012206}.

\begin{figure}[t]
\begin{indented}
\item[]
\begin{tabular}{cc}
\includegraphics[scale=.7,angle=0]{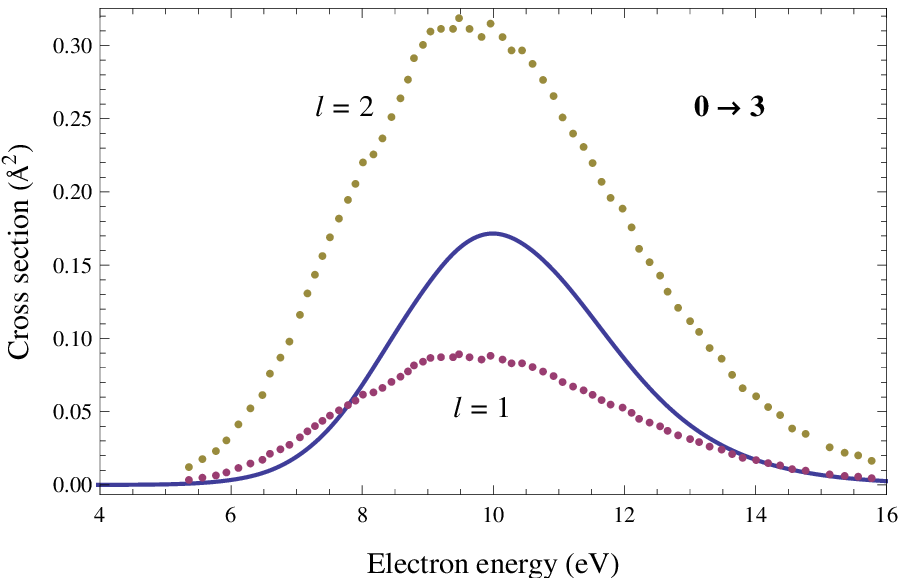}&
\includegraphics[scale=.7,angle=0]{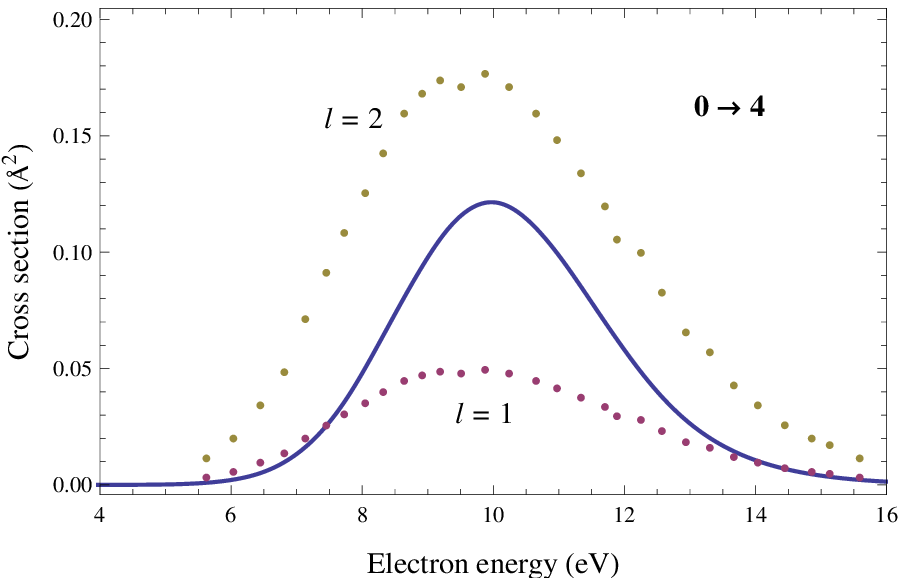}
\end{tabular}
\end{indented}
\caption{Computed cross sections (solid line) compared with the angular integrated Allan's results (dotted lines) assuming a pure $p$-wave ($l=1$) and a pure $d$-wave ($l=2$) in the outgoing wave function. \label{fig:cfrallan_l}}
\end{figure}

Returning to Fig.~\ref{fig:cfr_all}, there is an apparent discrepancy between the computed and experimental cross sections for peak height near 10~eV, in particular for the transitions $0\to 3$ and $0\to 4$. Actually the differential cross sections for these transitions were measured at a fixed angle and the integrated cross sections were obtained by assuming a predominant $p$-wave in the outgoing electron, which implies $l=1$, as suggested by the experimental results of Shyn \emph{et al.}~\cite{PhysRevA.48.1214}. The angular distribution analysis of differential cross sections made by Allan~\cite{allan_0953-4075-28-23-021} suggests, conversely, mixing between $p$- and $d$-wave in the outgoing wave function.  Fig.~\ref{fig:cfrallan_l} shows the differential cross sections of Allan  converted to integral cross sections assuming either a pure $p$-wave or a pure $d$-wave in the outgoing wave function. It can be seen that a $p-d$ mixture is compatible with the our results.

\begin{figure}[t]
\begin{tabular}{ccc}
\includegraphics[scale=.6,angle=0]{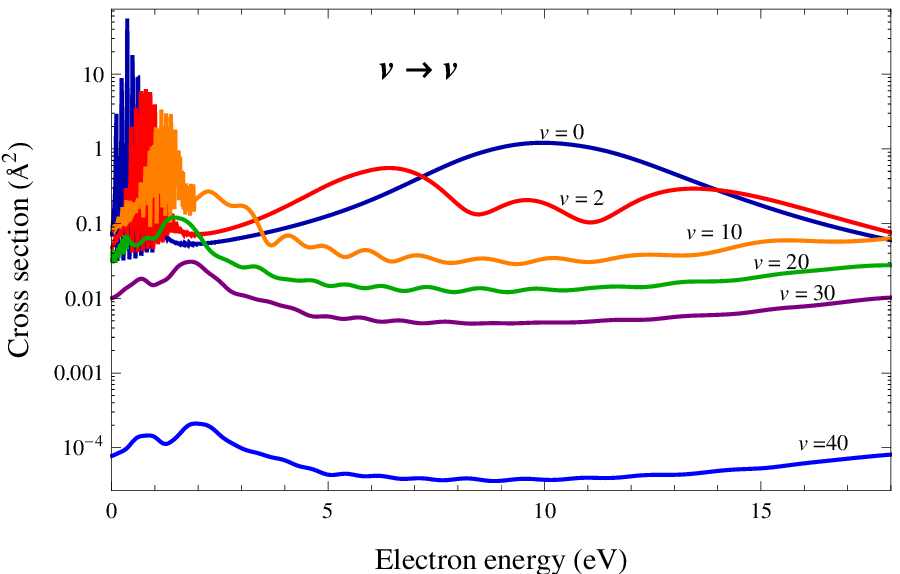}&
\includegraphics[scale=.6,angle=0]{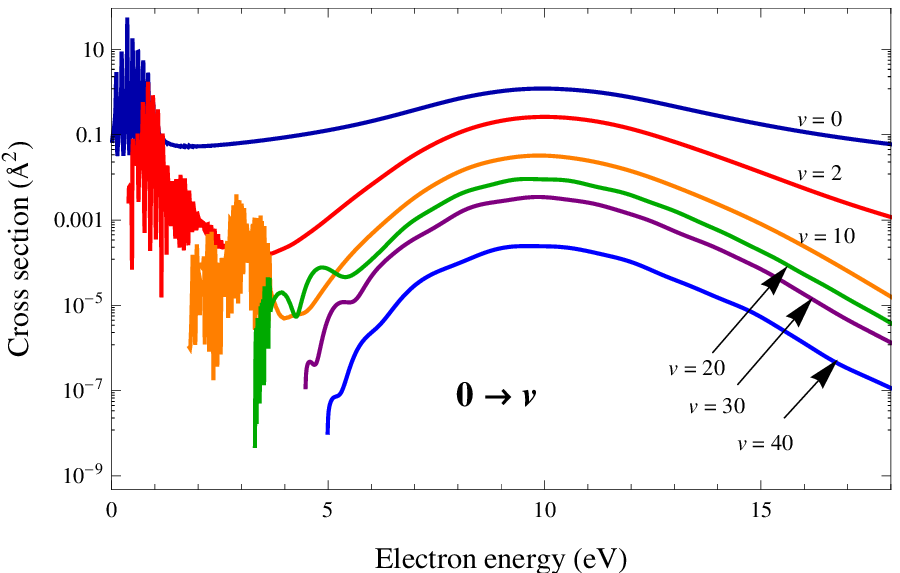}&
\includegraphics[scale=.6,angle=0]{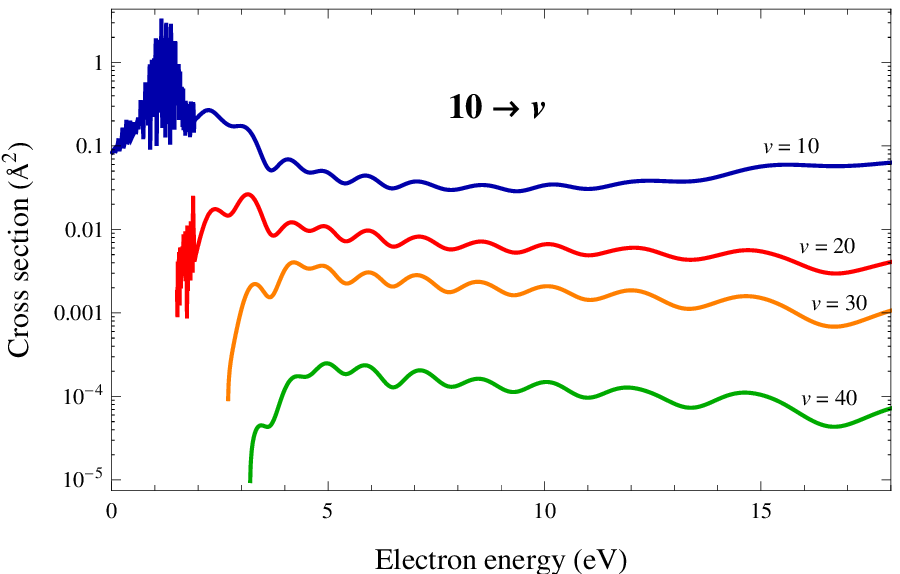}
\\
\includegraphics[scale=.6,angle=0]{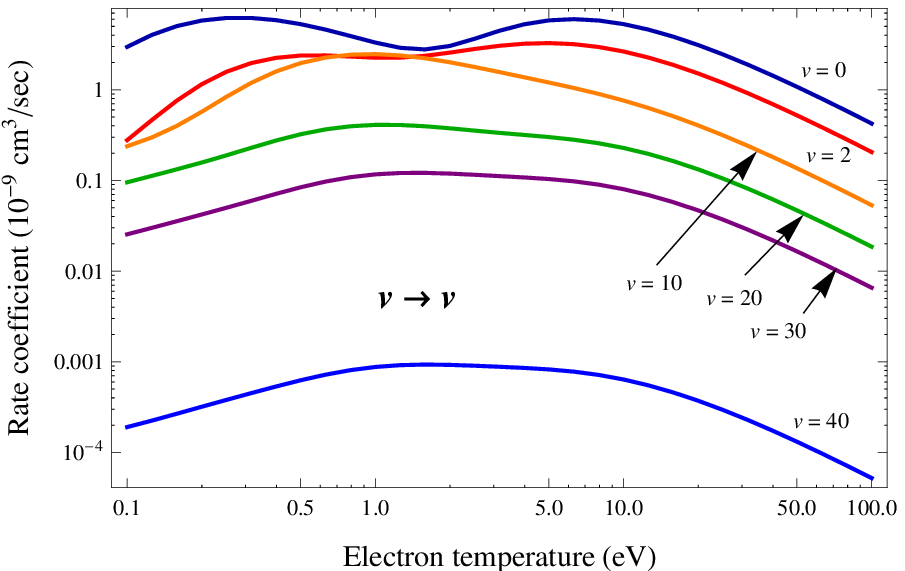}&
\includegraphics[scale=.6,angle=0]{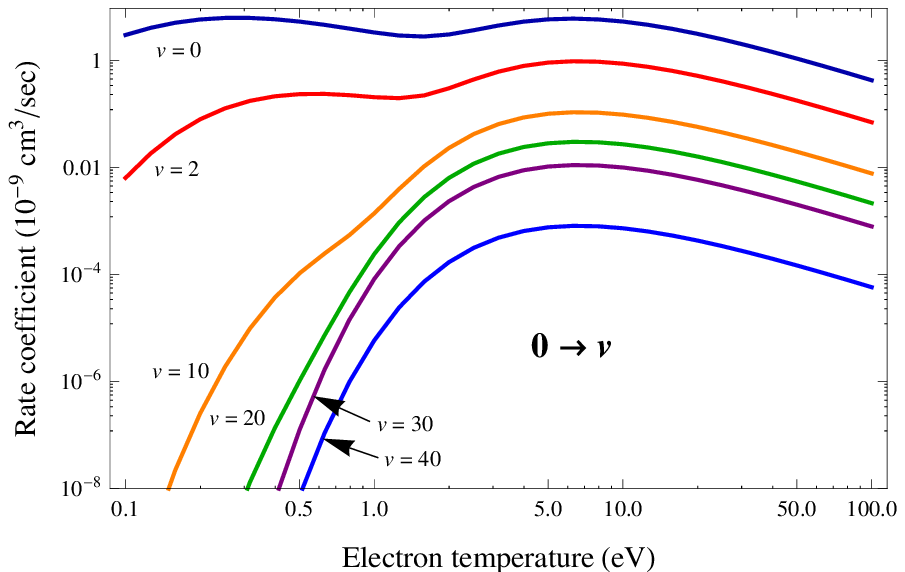}&
\includegraphics[scale=.6,angle=0]{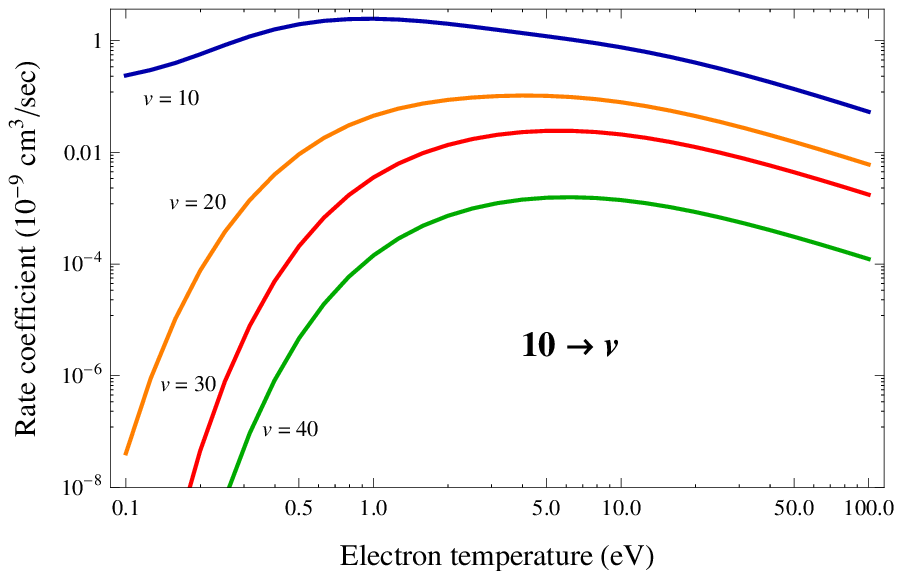}
\end{tabular}
\caption{Computed electron--O$_2$ vibrational-excitation cross sections (upper panels) and the corresponding rate coefficients (lower panels) for the transitions shown in the plots and for the rotational quantum number $J=1$. \label{fig:O2xsec_J0}}
\end{figure}

Figure~\ref{fig:O2xsec_J0} summarizes our new results for electron--O$_2$ RVE cross sections (top panels) and the corresponding rate coefficients (bottom panels) for the fundamental rotational quantum numbers $J=1$. Elastic and inelastic transitions are shown. Among the inelastic transitions, the most important in plasma kinetics, it can be seen that cross sections and rate coefficients decrease very rapidly at low energy and temperature as $\D v=v'-v$ increases. Conversely, in the 10~eV region, a more regular behavior is observed.

\begin{figure}[t]
\begin{indented}
\item[]
\begin{tabular}{cc}
\includegraphics[scale=.6,angle=0]{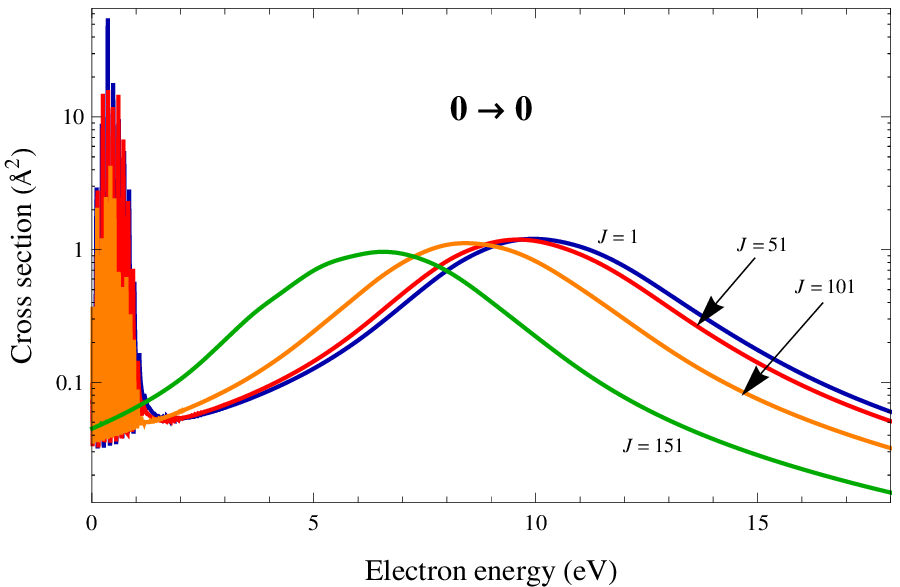}&
\includegraphics[scale=.6,angle=0]{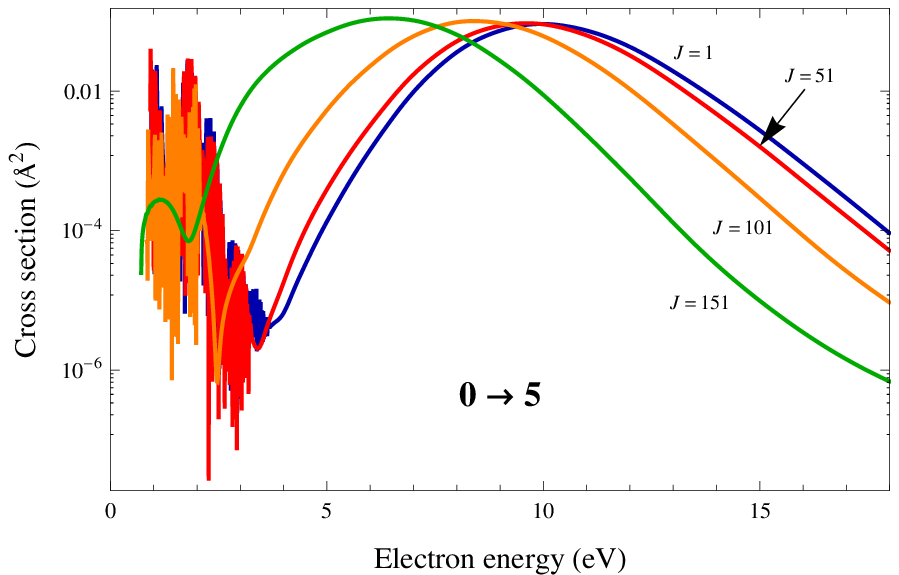}
\\
\includegraphics[scale=.6,angle=0]{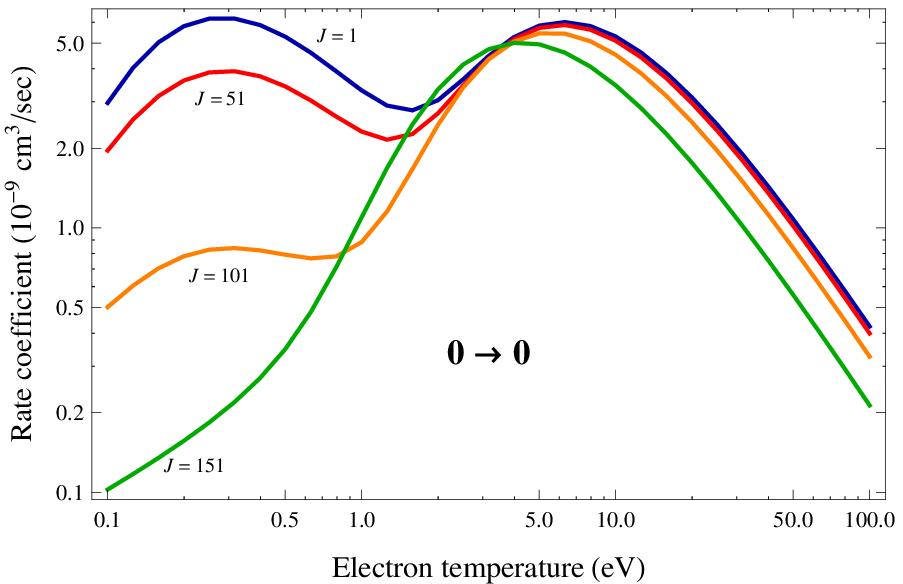}&
\includegraphics[scale=.6,angle=0]{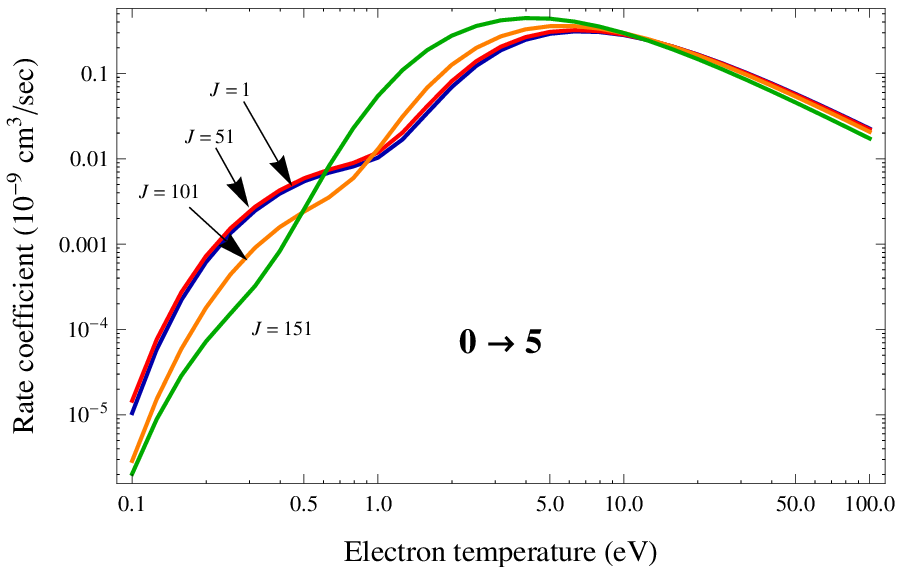}
\end{tabular}
\end{indented}
\caption{Effect of rotational quantum number $J$ on the electron--O$_2$ cross sections (upper panels) and on the corresponding rate coefficients (lower panels) for the vibrational excitations $0\to 0$ and $0\to5$. \label{fig:O2xsec_J}}
\end{figure}

The higher rotational quantum numbers are taken into account in Fig.~\ref{fig:O2xsec_J} where only the dominant transitions with $\Delta J=0$ are considered. In particular, the behavior of $\sigma_{00}$ and $\sigma_{05}$ are shown for $J=1$, 51, 101 and 151. The effect of increasing $J$ on the cross sections is to cut the peaks at low energy and to shift the position of the resonance at 10~eV toward lower energies, while the magnitude of the cross sections remains in fact unchanged. As a consequence, in a collisional-radiative kinetic description of a plasma containing molecular oxygen, rotational excited states of O$_2$ should not be neglected, in particular for low electron temperatures.

Finally, it can be observed also that the local maximum in the rate constant is related to that in the corresponding cross section. It occurs at the temperature: \be \tilde T_e = \frac{2\,\tilde\e}{3}\,,\ee where $\tilde\e$ is the electron energy corresponding at local maximum in cross section.

\section{Conclusions}

A complete set of resonant electron--O$_2$ vibrational-excitation cross sections and corresponding rate coefficients, in the range of energies below 4~eV and around 10~eV, were computed. Transitions between all the 42 vibrational levels of O$_2(\textrm{X}\ ^3\Sigma^-_g)$ for rotational quantum number $J=1$ were considered. The effects of rotational quantum number were also taken into account for $J=1\dots151$.

We confirm, at low-energy and low $J$, the dominance of $^2\Pi_g$ resonance state of O$_2^-$ for transitions involving lower vibrational levels but for high vibrational levels  we find a non-negligible contribution coming from the other resonances. At 10~eV  we find that the cross section is mainly determined by the $^4\S^-_u$ resonance state, but with an important contribution coming from the $^2\S_u^-$ and $^2\P_u$ states. Our results are in good agreement with the data available in literature. Finally, we find a non-negligible effect of rotation on the cross sections and rate constants, in particular for vibrationally inelastic transitions.

Our  next step will be the calculation of the dissociative electron attachment process  using the same input data. A  comparisons of these results with the available experimental data will be given elsewhere.

The full set of data obtained in the present work is available \emph{via} the website of the Phys4Entry project~\cite{F4Edatabase} and as supplementary material to this article.

\ack
The authors wish to thanks Dr. D.~Bruno (CNR-IMIP, Bari, Italy), Prof. S.~Longo (Universit\`{a} di Bari, Italy) and Dr. S.~Yurchenko (University College London, UK) for the careful reading of the manuscript. The research leading to these results has received funding from the European Community's Seventh Framework Programme (FP7/2007-2013) under grant agreement n$^\textrm{o}$ 242311.

\section*{References}
\addcontentsline{toc}{section}{References}

\bibliographystyle{is-unsrt}{}  %  <----
%\bibliographystyle{pccp}{}
%\bibliographystyle{rsc}{}
%\bibliographystyle{ieeetr}{}
%\bibliographystyle{abbrv}{}
%\bibliographystyle{apsrev4-1}{}

%\bibliography{C:/Users/vincenzo/Desktop/appunti/resonant}

\end{document}